\documentstyle[preprint,aps,eqsecnum]{revtex}
\begin{document}
\draft
\newcommand{\comm}[1]{\underline{\tt #1}}

\tightenlines
\newcommand{\modegy}{\,({\rm od}\,1)}
\newcommand{\slk}{$sl(2)_{-1}$}
\newcommand{\modketto}{\,({\rm mod}\,2)}
\newcommand{\modpi}{\,({\rm mod}\,\pi)}
\newcommand{\modketpi}{\,({\rm mod}\,2\pi)}
\newcommand{\J}{{\cal J}}
\newcommand{\ch}{{\rm ch}}
\newcommand{\sh}{{\rm sh}}
\newcommand{\th}{{\rm th}}
\newcommand{\sn}{{\rm sn}}
\newcommand{\cn}{{\rm cn}}
\newcommand{\dn}{{\rm dn}}
\newcommand{\am}{{\rm am}\!}
\newcommand{\ordo}[1]{{\cal O}(#1)}
\newcommand{\bml}{\begin{mathletters}}
\newcommand{\eml}{\end{mathletters}}
\newcommand{\be}{\begin{equation}}
\newcommand{\ee}{\end{equation}}
\newcommand{\ba}{\begin{array}}
\newcommand{\ea}{\end{array}}
\newcommand{\bea}{\begin{eqnarray}}
\newcommand{\eea}{\end{eqnarray}}
\newcommand{\bfl}{\begin{flushleft}}
\newcommand{\efl}{\end{flushleft}}
\newcommand{\nn}{\nonumber}
\newcommand{\smi}{\!-\!}
\newcommand{\spl}{\!+\!}

\def\eqalign#1{
\null \,\vcenter {\openup \jot \ialign {\strut \hfil $\displaystyle {
##}$&$\displaystyle {{}##}$\hfil \crcr #1\crcr }}\,}

%%%%%%%%%%%%%%%%%%%%%%%%%%%%%%%%%%%%%%%%%%%%%%%%%%%%%%%%%%%%%%%%%%
\pagestyle{myheadings}
\title{
{\flushright{\small MIT-CTP-2816\\cond-mat/9901184\\} \vspace{.2in}}
Scaling limit of the one-dimensional XXZ Heisenberg chain\\  
with easy axis anisotropy}
\author{\bf\large Tam\'as Hauer}
\address{Center for Theoretical Physics,\\
LNS and Department of Physics\\
Massachusetts Institute of Technology\\
Cambridge, Massachusetts 02139, U.S.A.}
\author{\bf\large Attila R\'akos, Ferenc Woynarovich}
\address{Institute for Solid State Physics\\
Hungarian Academy of Sciences\\
1525 Budapest 114, Pf 49.}
\maketitle
\begin{abstract}
We construct the scaling limit of the easy axis XXZ chain.  This limit
is a subtle combination of approaching the isotropic point, and
letting the lattice spacing to zero to obtain a continuous model with
a finite mass gap.  We give the energy difference between the two
lowest energy states (the two `vacua') and analyze the structure of
the excitation spectrum of the limiting model. We find, that the
excitations form two sets corresponding to the two vacua.  In both
sets the dressed particles are described by Bethe Ansatz like
equations (higher level Bethe Ansatz), and the two sets can be
distinguished through a parameter entering into these secular
equations.  The degenerations in the spectrum can be interpreted as
originating from an SU(2) symmetry of the dressed particles. 
The two particle scattering matrices obtained from the secular equations are
consistent with this symmetry, and they differ in an overall sign in
the two sectors.
\end{abstract}


%%%%%%%%%%%%%%%%%%%%%%%%%%%%%%%%%%%%%%%%%%%%%%%%%%%%%%%%%%%%%%%%%%
\setlength{\parskip}{2ex}
\setlength{\parindent}{0em}
\setlength{\baselineskip}{3ex}

\section{Introduction}

The XXZ Heisenberg chain is defined by the Hamiltonian 
\be
\label{Ham} 
H_{XXZ}={V\over a}\sum_{n=1}^N
\left(\left(S_n^xS_{n+1}^x+S_n^yS_{n+1}^y\right)+
\rho\left(S_n^zS_{n+1}^z-1/4\right)\right)\,. 
\ee 
The Hilbert space is the tensor product of $N$ spaces furnishing the
doublet representation of SU(2). $S_n^{x,y,z}$ are the spin operators
acting on the $n$th site and in (\ref{Ham}) the $(N\!\!+\!\!1)$th site
is identified with the first one. We consider the easy axis region
defined by
\be 
\rho=\cosh\gamma>1\,.
\ee 
The $V/a$ factor with $a$ being the lattice spacing and $V=2/\pi$ is
included for the sake of proper normalization.  

The Hamiltonian (\ref{Ham}) can be diagonalized by Bethe Ansatz (BA)
\cite{Be,Or,desClGa,Yang}, the BA equations (BAE) were analyzed in
\cite{BVV,ViWo,deVW}.  Due to these studies it is well known by now, that
the chain with even number of sites has two states in which there are
no free parameters \cite{ViWo}. These are the ground state and the
first excited state, and as the difference in their energy disappears
exponentially fast with the length of the chain $N$ \cite{deVW,Bax},
they are often referred to as the two ground states, although it is more
precise to call them the two vacua.  The other
excited states are described in terms of a few parameters characterizing
some kind of dressed particles. These
posses a gap, and their parameters satisfy BA type equations, the so
called higher level Bethe Ansatz equations (HLBAE) \cite{BVV,ViWo}. The
excited states form two groups which can be associated with the two
ground states \cite{ViWo}. 

The model has two relativistic continuum limits \cite{McCWu}.  
One is constructed by setting $\gamma=0$ first
and taking the $a\to0$ (together with $N\to\infty$ but $Na=L$)
continuum limit afterwards resulting a massless SU(2) conformal field
theory. To obtain the other relativistic limit one has to perform the
$\gamma\to0$ and $a\to0$ simultaneously keeping the gap in the
properly normalized spectrum constant. The aim of the present work is
to study the details of this second limit, and analyze the limiting
theory. Our results are as follows.

\begin{itemize}

\item The limiting theory is a massive relativistic theory with a mass
$M$, if the continuum limit is performed in such a way, that
\be\label{rellim}
\gamma\to0\,,\quad\quad a\to0\,,\quad\quad
{4\over a}\exp\left\{-{\pi^2\over2\gamma}\right\}\to M\,.
\ee
As the length of the chain $L=Na$ is kept finite, also the number of
sites must be adjusted:
\be\label{rellim'} 
N={LM\over4}\exp\left\{{\pi^2\over2\gamma}\right\}\to \infty\,.
\ee
Our expression for the mass $M$ differs from that found in \cite{McCWu}
in a prefactor of the exponential.
\item In the above limit the difference in the energies of the two
`ground states' (physical vacua) stays finite but exponentially small
in the length of the system:
\be\label{endifiscl}
\Delta E_0=\sqrt{{8M\over\pi L}}e^{-LM}\,.
\ee
\item The excited states of the system can be described in terms of
excitations (dressed particles). Each particle is characterized by a
rapidity we denote by $\vartheta$. The energy and momentum
contribution of the particles are the sums of the contributions of the
individual particles
\be
E-E_0=\sum\varepsilon(\vartheta)\,,\quad\quad P=\sum p(\vartheta)\,,
\ee
with
\be
\varepsilon(\vartheta)=M\cosh\vartheta\,,\quad\quad 
p(\vartheta)=M\sinh\vartheta\,.
\ee
The rapidities of the particles must satisfy a set of BA type
equations:
\bml\label{ge}
\bea\label{gea}
Lp(\vartheta_h)&=&2\pi({\cal I}_h+I_0)-\sum_l^{n(\vartheta)}
\phi\left({\vartheta_h-\vartheta_l\over\pi}\right)
+\sum_{\alpha}^{n(\kappa)}
2\tan^{-1}{\vartheta_h-\kappa_{\alpha}\over\pi/2}\,,\nn\\  
{\cal I}_h&=&{1\over2}\left({n(\vartheta)-2n(\kappa)\over2}
\right)\modegy\,,\quad\quad I_0=0\ {\rm or}\ 1/2\,,
\eea
and
\bea\label{geb}
\sum_h^{n(\vartheta)}2\tan^{-1}{\kappa_{\alpha}-
\vartheta_h\over\pi/2}
&=&2\pi{\cal J}_{\alpha}+\sum_{\beta}^{n(\kappa)}
2\tan^{-1}{\kappa_{\alpha}-\kappa_{\beta}\over\pi}\,,\nn\\
{\cal J}_{\alpha}&=&\left({n(\kappa)-n(\vartheta)+1\over2}
\right)\modegy\,.
\eea
\eml
Here $n(\vartheta)$ is the number of particles, 
the set of variables $\kappa$ is needed to describe the internal
symmetry of the states, their number $n(\kappa)$ obeys
$n(\vartheta)-2n(\kappa)\geq0$, and
\be\label{phifv}
\phi(x)={1\over i}{\rm ln}{\Gamma\left({1\over2}-i{x\over2}\right)
                          \Gamma\left(1+i{x\over2}\right)\over
                          \Gamma\left({1\over2}+i{x\over2}\right)
                          \Gamma\left(1-i{x\over2}\right)}\ .
\ee

\item The excited states form SU(2) multiplets in which the energy is the same.
Each {\em multiplet}
is characterized by one solution of the equations (\ref{ge}). The
number of states belonging to one multiplet is
$(n(\vartheta)-2n(\kappa))+1$.  Within a multiplet the states are
labeled by the value of the $S^z$ taking the values  
$n(\vartheta)/2\smi n(\kappa),
\ n(\vartheta)/2\smi n(\kappa)\smi1,\ \ldots,\ -n(\vartheta)/2\spl n(\kappa)$. 
This indicates, that the particles have SU(2) symmetry 
or a symmetry producing the same multiplet structure,
and the $z$ component of the spin connected to this symmetry coincides with 
the $z$ component of the real spins building up the original chain.
\item The solutions of (\ref{ge}) form two groups, one for $I_0=0$ and
one for $I_0=1/2$. We argue, that the two sets of excited states are
the excitations of the two vacua. In both sets the $S^z=0$
states, including the vacua, are eigenstates of flipping all spins, but with
different eigenvalues. In the case of the two particle
excitations the symmetry of the singlets is the same as that of the 
corresponding vacuum, 
and that of the triplets is the opposite. This is the same structure
as found in the $XXX$ chain, where it is a consequence of the SU(2) symmetry
of the Hamiltonian (and this structure can be interpreted in terms of SU(2)
with a modified coproduct, or equivalently in terms of a $q$-deformed 
SU(2) at $q=-1$).
\item The two particle scattering matrix can be given
up to an overall phase as
\be\label{smm1}
\hat S(\Delta\vartheta)=-\exp\left\{i\left(2\pi I_0+
\phi\left({\Delta\vartheta\over\pi}\right)
\right)\right\}
\left(\hat{P}_{tr}+{\Delta\vartheta+
  i\pi\over \Delta\vartheta-i\pi}\hat{P}_{s}\right)\,,
\ee
where $\hat{P}_{tr}$ and $\hat{P}_{s}$ are the 
projectors on the triplet resp.~singlet subspace of the two spins.
\end{itemize}

All these shows, that the scaling limit (SL) of the XXZ yields a massive 
relativistic theory (in duplicate) which is of the same structure as 
the massive sector of the theory obtained through the SL of the attractive 
Hubbard chain and identified as a regularization of the SU(2) symmetric
chiral invariant Gross-Neveu (CGN) model \cite{WoFo1,WoFo2}.  
It is widely accepted, that the XYZ chain is a lattice regularization
of the sine-Gordon (SG) theory or the massive Thirring
model (MTM) \cite{Lut} (as these latter two are equivalent \cite{Col}).
It is also known, that the MTM/SG theory at a special value of the 
coupling is equivalent to the CGN model (up to a free massless boson field),
more over this value of the coupling corresponds to the antiferromagnetic
(our $\rho$=1) point of the XYZ chain. This way recovering the massive 
sector of the CGN model is not unexpected. It is remarkable, however, that
the SL of the XXZ chain is taken through couplings not corresponding
to stable MTM/SG theories. Our detailed analysis also rises some        
questions connected with the mass-formula  and the symmetry of the vacua.

The (\ref{rellim}) mass-formula is not of the form found in the case of
the Hubbard chain
(a prefactor $\sqrt{\gamma}$ is `missing'), 
and it defines a $\beta$-function different from the
one expected based on perturbation calculations. This may be connected 
with the fact, that the $\rho$=1 point in the space of the couplings 
of the XYZ model is singular (in the sense, that certain quantities 
like the ground-state energy of the XXZ chain are singular in this point
\cite{desClGa}), thus approaching this point from different directions
may lead to different results. In our SL the isotropic point is approached 
along the XXZ line while the limit taken in \cite{Lut} involves the 
XY-type anisotropy.

Another interesting point is the existence of two very similar but not
completely identical limiting theories. The two theories are distinguished
by the symmetry of the vacuum with respect to reversing all spins. One has to 
note, however, that measuring this symmetry in the continuum 
limit may encounter difficulties, as the SL of the corresponding operator
can not be constructed directly. This symmetry has not been studied in 
other representations of the CGN model although it 
may be present also in other cases. In the SL of the half-filled 
Hubbard chain it is even stronger in the sense that the vacuum is a 
singlet of both the spins and the isospins, and it is symmetric under 
both the spin and isospin reversal.    

The paper is organized as follows. In Section \ref{sec:BA} we summarize 
those properties of the XXZ chain we need to construct its scaling limit
and we also propose a formula to calculate the symmetry with respect
to flipping all spins. 
In Section \ref{sec:skalalim} the SL is constructed and the 
properties of the limiting theory are analyzed. Specially we construct the
limiting process resulting in a relativistic dispersion 
(\ref{sec:spectr}),
analyze the vacua in this limit
(\ref{sec:vakumok}),
derive the secular equations of the limiting theory
(\ref{sec:seceq}),
we give the SU(2) multiplet structure of the eigenstates
(\ref{sec:multipl}),
derive the two particle $S$-matrix
(\ref{sec:Smatrix}),
we study the reflection symmetry of the two particle $S^z=0$ states 
(\ref{sec:reflsim}), and we also argue,
that the boundary condition obeyed by the particles of the limiting 
theory is connected to the parity properties of $N$ kept unchanged in the
$N\to\infty$ limit 
(\ref{sec:boundary}).
The more technical details are collected in appendices. The properties of the
elliptic functions we use are listed in Appendix \ref{sec:lista}. The 
behavior of the two vacua are calculated in Appendix \ref{sec:alapok},
and in Appendix \ref{sec:strstr} we check if the the approximations used
to derive the HLBAE of the lattice model remain valid also in the SL. In Appendix
\ref{sec:solstr} we describe the structure of the solutions of the HLBAE
and give the classes of solutions which become degenerate in the SL.
Finally some details of our numerical calculations are given in 
Appendix \ref{sec:numerics}.

%%%%%%%%%%%%%%%%%%%%%%%%%%%%%%%%%%%%%%%%%%%%%%%%%%%%%%%%%%%%%%%%%%%
\section{Bethe Ansatz solution of the XXZ chain}
\label{sec:BA}

\subsection{The BA equations}

This section is intended to summarize the BA solution of the XXZ chain 
\cite{BVV,ViWo,deVW} with emphasis on those points which are relevant
to the construction of the SL. 

First let us recall some properties of the Hamiltonian (\ref{Ham}).
In the literature two versions of the XXZ model are commonly found
which are distinguished by the relative sign of the
$(S_n^xS_{n+1}^x+S_n^yS_{n+1}^y)$ and $(S_n^zS_{n+1}^z)$ terms. The
two theories are equivalent if $N$ is even, the corresponding
transformation on the Hilbert spaces is generated by
$\prod_{k}\sigma^z_{2k}$ which relates the spin operators as:
$\{S_{2k}^x,S_{2k}^y,S_{2k}^z\}\rightarrow
\{-S_{2k}^x,-S_{2k}^y,S_{2k}^z\}$. As this is a symmetry of the
Hamiltonian, the energy spectrum does not change, but it does have an
effect on the eigenvalues of the momentum operator which is relevant
to the connection to the continuum limit. The momentum of every spin
wave is shifted by $\pi$ which affects all the states with an odd
number of spin waves, including the ground state whenever $N/2$ is
an odd integer. There is another useful operator,
which commutes with the Hamiltonian, but whose
eigenvalues are affected: 
\be\label{tukor}
\hat\Sigma=\prod_{n=1}^N\,\sigma_n^x\,, 
\ee 
which represents a reflection on the $x$-axis.
(In a basis given by the products of the $S^z$ eigenstates of the individual 
spins $\hat\Sigma$ simply flips all the spins (the up ones down and 
the down ones up)). The symmetry connected with this operation we call 
spin reversal or reflection symmetry.) 
The $S^z=0$ BA eigenstates of (\ref{Ham}) are expected to be 
also eigenstates of
(\ref{tukor}). For the sake of definiteness we have chosen the
positive sign in (\ref{Ham}) and note that when $N$ is
an integer multiple of four, the results should agree with the ones
obtained from the opposite convention. 

The Hamiltonian (\ref{Ham}) and $S^z$ can be simultaneously
diagonalized by the Bethe Ansatz (BA) and due to the symmetry
corresponding to inverting the spins it is sufficient to consider only
$S^z\geq 0$ states. The eigenvectors are explicitly given in terms of
a wave number set $\left\vert\{k_1\ldots k_r\}\right\rangle$, for the
states with $S^z=N/2-r$, ($r\leq{N/2}$)
\be 
\label{BAstate}
\left\vert\{k_1\ldots k_r\}\right\rangle= 
\sum_{n_1<\ldots<n_r} \left( \sum_{\cal P}
\exp\left\{i\sum_\alpha k_{{\cal P}_\alpha} n_\alpha  
+ {i\over2}\sum_{\alpha<\beta} 
\Psi_{{\cal P}_\alpha {\cal P}_\beta}\right\}\prod_{n_\alpha}\sigma_{n_\alpha}^- 
\left\vert F\right\rangle\right),
\ee
where the summation is over all permutations $\cal P$ of $\{1,\ldots,
r\}$ and the wave numbers $k_\alpha$ together with the phase shifts 
$\Psi_{\alpha\beta}$ satisfy the equations
\be
\label{fazis}
\cot{\Psi_{\alpha,\beta}\over2}=-\rho
{\cot{k_{\alpha}\over2}-\cot{k_{\beta}\over2}\over
(1-\rho)\cot{k_{\alpha}\over2}\cot{k_{\beta}\over2}-(1+\rho)}\,, 
\ee
\be\label{bea1}
\exp\left\{Nk_{\alpha}\right\}= \exp\left\{\sum_{\beta(\not=\alpha)}
\Psi_{\alpha,\beta}\right\}\,.
\ee
The energy and the momentum of the state are
\be\label{em}
E={V\over a}\sum_{\alpha}\left(\cos k_{\alpha}-\rho\right)\,;
\quad\quad Q=\sum_{\alpha} k_{\alpha}\,.
\ee
The equation (\ref{fazis}) can be solved for the phase shifts by
parameterizing $k_\alpha$'s in terms of new variables, $v_\alpha$
\cite{Yang} which are commonly called rapidities: 
\be
k_{\alpha}=\Phi(v_{\alpha},\gamma/2)\,,\quad\quad
\Psi_{\alpha,\beta}=\Phi(v_{\alpha}-v_{\beta},\gamma)\,,
\ee
with
\be
\Phi(z,\delta) \equiv {1\over i}
\ln{\sin(z+i\delta)\over\sin(z-i\delta)}\,.
\ee
When expressing (\ref{bea1}) in terms of the rapidities it turns out
to be useful to take the logarithm of the equation. In order to do
this one has to define the branch cuts of $\Phi(z,\delta)$. We
choose $\Phi(z,\delta)$
to be a continuous function of $z$ with
$\Phi(0,\delta)=\pi$ in the strip $\vert{\rm Im}\,z\vert<\delta$,
while in the regions  ${\rm Im}\,z>\delta$ and ${\rm Im}\,z<-\delta$
we choose those levels in which $\Phi\to\mp2i\delta$ 
if ${\rm Im}\,z\to\pm\infty$,
respectively. This results in cuts running along the lines 
${\rm Im}z=\pm\delta$,
${\rm Re}z\leq0$ and ${\rm Re}z\geq\pi$, and corresponds to choosing 
$f_1=f_2=0$ in \cite{ViWo}. 
Taking the logarithm of (\ref{bea1}) it becomes:
\be\label{BAe}
N\Phi(v_{\alpha},\gamma/2)=2\pi I_{\alpha}+
\sum_{\beta}\Phi(v_{\alpha}-v_{\beta},\gamma)\,,
\ee
where the $I_{\alpha}$ are half-odd integers and turn out to be very
useful quantum numbers for characterizing the states with macroscopic
number of spin waves \cite{Yang}.  The energy
and the momentum in terms of the rapidities are
\be\label{emphi}
E={V\over a}{\sinh\gamma\over2}\sum_{\alpha}
\Phi^{\prime}(v_{\alpha},\gamma/2)\,,
\quad\quad 
Q=\sum_{\alpha}\Phi(v_{\alpha},\gamma/2)\,.
\ee 
For technical reasons it is 
useful to restrict the real part of the rapidities to an interval of length 
$\pi$. We choose $0<{\rm Re}\,v_{\alpha}\leq\pi$, but note, that for our 
results it is important only, that the interval contains the point $\pi/2$ in
its interior. Note also, that the wave function, the energy and the momentum are
independent of the particular choice of this interval and 
branch cuts of $\Phi(z,\delta)$ (the latter modulo $2\pi$).

\subsection{The ground states}

In the ground state(s) of the XXZ model with $N$ even, there are $N/2$
rapidities ($S^z\!=\!0$), all of them are real and all the $I_j$
quantum numbers are {\em consecutive} half-odd integers satisfying
$I_{j+1}-I_j=-1$ for $\lambda_{j+1}>\lambda_j$ \cite{Yang}.  (Here and
in the following we denote the real rapidities by $\lambda_j$.)  For
large $N$ the distribution of $\lambda$'s can be well approximated by
a smooth density $\sigma_0(\lambda)$ satisfying the linear integral
equation \cite{desClGa,Yang,BVV,ViWo}:
\begin{eqnarray}
\label{BAinte}
-\Phi^{\prime}(\lambda,\gamma/2)&=&\sigma_0(\lambda)-{1\over2\pi}
\int\limits_{\Lambda}^{\Lambda+\pi}
\Phi^{\prime}(\lambda-\lambda^{\prime},\gamma)
\sigma_0(\lambda^{\prime})d\lambda^{\prime}\,, \\
\label{BAeLambda}
\mbox{with} \;\;\;\;\;\; N\Phi(\Lambda,\gamma/2)
&=&2\pi\left(I_1+1/2\right)+
\sum_{j}^{N/2}\Phi({\Lambda}-\lambda_{j},\gamma)\,.
\end{eqnarray}
Eq.(\ref{BAinte}) can be solved by Fourier
transformation leading to
\be\label{sigma0}
\sigma_0(\lambda)={K\over\pi^2}\,\dn\!
\left({2K\over\pi}\lambda,k\right)\;\;\;\;\;
\mbox{with} \;\;\;\;\; {K^{\prime}\over K}={\gamma\over\pi}\,,
\ee
where $\dn(w)$ is the Jacobian elliptic function and $K$ is the complete 
elliptic integral of the first kind with modulus $k$. (See Appendix 
\ref{sec:lista} for notations and some properties of the elliptic
functions.) 

By means of this density the sum over $j$ in (\ref{BAeLambda}) can be
calculated, and the rapidity set can be reconstructed. One finds,
that there are {\em two nonequivalent} sets of $N/2$ real
$\lambda_j$'s satisfying (\ref{BAe}) with $I_{j+1}-I_j=-1$
\cite{ViWo}.  These are:
\bml
\be
\lambda_j={\pi\over2K}F\left({2\pi\over N}(j-I_0),k\right)\,,
\quad I_j=2I_0-1/2-j\,,
\quad j=1,2,\ldots N/2
\ee
for $N/2=$even and
\be
\lambda_j={\pi\over2K}F\left({2\pi\over N}(j-1/2+I_0)),k\right)\,,
\quad I_j=1/2-2I_0-j\,,
\quad j=1,2,\ldots N/2
\ee
\eml 
for $N/2=$odd,
with $F(w,k)$ being the elliptic integral of first kind. The two sets
are distinguished by $I_0$ taking the values of $1/2$ and $0$,
respectively. These two states of lowest energy are almost degenerate.
The true ground state corresponds to $I_0=1/2$ and the difference
between their energy is exponentially small in the length of the
chain \cite{Bax,deVW}. The leading term is:
\be\label{endifi}
\Delta E_0=E(I_0\!=\!0)-E(I_0\!=\!1/2)=
{V\over a}{\sqrt{8k^{\prime}}\over\pi^{3/2}}\sh\gamma 
K{k_1^{N/2}\over N^{1/2}}\ ,
\ee
where $k'=\sqrt{1-k^2}$ and $\sqrt{k_1}=(1-k')/k$. As $\Delta E_0$ is
exponentially small in the length of the chain and -- as we shall see
-- all the other states are separated from the lowest two by a finite
gap, we call these the two ground states or the physical vacua.  The
two vacua can be distinguished for instance by their momentum, $Q$ and
the eigenvalue, $\Sigma$ of the operator (\ref{tukor}) \cite{deVW}:
\bea
\label{pi0}
Q(I_0,N)&=&2\pi(I_0-1/2+N/4)\modketpi \\
\label{tukorertek}
\Sigma(I_0,N)&=&(-1)^{(2I_0-1+N/2)}\,.
\eea

\subsection{The excited states}

The states other than the two vacua are called excited states and the
ones of not too high energy can be described in terms of some dressed 
particles. These are created by leaving holes in the sequence of the
$\lambda_j$'s, and by introducing complex rapidities. The smooth
density $\sigma(\lambda)$ of the real rapidities is still a good 
approximation but 
(\ref{BAinte}) receives contributions from both the holes and the
complex rapidities. When one solves for $\sigma(\lambda)$ we are
left with a few equations relating the parameters of the excitations
only \cite{BVV,ViWo}:
\bea
\label{hlbae1a}
1\!&=&
\exp i\!\left\{\!N\!\left(\!\am\left({2K\over\pi}\theta_h,k\!\right)
\smi{\pi\over2}\right) \smi
2\pi I_0  \spl
\sum_{b=1}^{n(\theta)}\left(\!{\cal F}
(\theta_h\smi\theta_b,\gamma)\smi\frac{\pi}{2}\right) \spl
\sum_{\alpha}^{n(\chi)}
\Phi\!\left(\!\theta_h\smi\chi_{\alpha},\frac{\gamma}{2}\right)\!\right\}\! \\
\label{hlbae1b}
1\!&=&
\exp i\!\left\{\sum_{h=1}^{n(\theta)}
\Phi(\chi_{\alpha}-\theta_h,\gamma/2) - 
\sum_{\beta=1}^{n(\chi)}\Phi(\chi_{\alpha}-\chi_{\beta},\gamma) - \pi
\right\} 
\eea
where 
\be
{\cal F}(x,\gamma)\equiv x+\sum_{m=1}^{\infty}{e^{-\gamma m}\over
m\cosh(\gamma m)}\sin(2mx)\,,
\ee
and $I_0$ is again either $1/2$ or 0.
The positions of the holes in the real-rapidity distribution are
denoted by $\theta_h$ and their number is denoted by $n(\theta)$,
whose parity is the same as that of $N$. 
The variables $\chi$ represent the set of complex rapidities $z$ in the
following way.
If for a $\chi$ $\vert{\rm Im}\chi\vert<\gamma/2$, it represents a close
pair  
\bml\label{komplexek}
\be\label{kozeli}
z^{\pm}=\chi\pm i\gamma/2+\ordo{e^{-N\Omega(z^+,\gamma)}}
\ee
with $\Omega(z,\gamma)$ being a positive valued function of the order
of unity (as long as $\gamma>0$). If $\vert{\rm Im}\chi\vert>\gamma/2$,
the $\chi$ represents a wide root: 
\be\label{tavoli}
z=\chi+i\,{\rm sgn}({\rm Im}\chi)\gamma/2\,.
\ee
\eml
As the $\chi$s are either real, or form complex conjugated pairs, the typical
complex rapidity configurations are the 2-strings (the two $z$ represented
by a real $\chi$),
the quartets (the four complex rapidities represented
by a complex conjugated pair with $\vert{\rm Im}\chi\vert<\gamma/2$), and 
wide pairs (corresponding to a complex conjugated pair of $\chi$s with
$\vert{\rm Im}\chi\vert>\gamma/2$). The
total number of the complex parameters $\chi_{\alpha}$ is denoted by $n(\chi)$.

The energy and the momentum of the excited states can be expressed in
terms of the parameters of the holes only and are given as:
\bea
\label{epstheta}
E=E_0+\sum_h\epsilon(\theta_h) &\;\;\;\;\mbox{with}\;\;\;\;& 
\epsilon(\theta)={V\over a}{K\over\pi}\sinh\gamma
\,\dn\left({2K\over\pi}\theta,k\right)\,, 
\\ \label{ptheta}
Q=Q(I_0,N)+\sum_h q(\theta_h) &\;\;\;\;\mbox{with}\;\;\;\;& 
q(\theta)=\am\left({2K\over\pi}\theta,k\right)-{1\over2}\pi,
\eea
with $E_0$ being the ground state energy, and $Q(I_0,N)$ is given in
(\ref{pi0}). The appearance of $Q(I_0,N)$ in the momentum suggests, that
the $I_0=1/2$ resp.\ $I_0=0$ excited states should be considered as
excitations above the $I_0=1/2$ resp.\ $I_0=0$ vacua.
The relation between the functions $\epsilon(\theta)$ and $q(\theta)$
give the dispersion relation of the particles:
\be\label{ep}
\epsilon(q)={V\over a}{K\over\pi}\sinh\gamma\sqrt{1-k^2\cos^2q}\,.
\ee
The spin of the state characterized by a solution is 
\be\label{spin}
S^z={1\over2}n(\theta)-n(\chi)\,.
\ee
The states with $S^z<0$ are obtained by flipping all spins, for
example using the operator $\hat\Sigma$ of (\ref{tukor}). The $S^z=0$ BA 
eigenstates are expected to be eigenstates of this operation: otherwise
certain points of the spectrum were twofold degenerated (with the two states
connected by $\hat\Sigma$).
This can happen {\em accidentally} but is not forced by the symmetry, as   
$\hat\Sigma$ has one-dimensional representations only. 
Recently Doikou and Nepomechi has proposed a formula for the parity
of the $S^z=0$ in the planar ($\rho<1$) region \cite{DoNe}.
Now we {\em conjecture}, that in the easy axis region we study, 
the spin reversal symmetry of the $S^z=0$ states in analogy with 
their formula is given as
\be\label{sumall}
\Sigma=(-1)^{\mu},\quad\quad{\rm with}\quad\quad
\mu=\frac{N}{2}+\frac{2}{\pi}\sum^{N/2}_{\alpha=1}v_{\alpha}\,.
\ee
We have not proved this formula, but performed numerical calculations 
to see this symmetry
(Appendix \ref{sec:numerics}): solving the BA
equations numerically for a number of $S^z=0$ states has shown, that the
wave function is either symmetric or antisymmetric under 
$\hat\Sigma$, and the value of $\Sigma$ is correctly given by (\ref{sumall}).
In the large $N$ limit (\ref{sumall}) can be transformed into a simpler
form using the density of the real roots \cite{ViWo}:
\be\label{simsum}
\mu=(2I_0+1)+\frac{N}{2}+\frac{2}{\pi}\left\{
\sum_{\alpha}^{n(\chi)}\chi_{\alpha}-\frac{1}{2}\sum_h^{n(\theta)}\theta_h
\right\}\modketto.
\ee

In summary, the structure of the states can be described as follows:
the system has two vacua found in the $N$ even system, one of them is
symmetric, the other is anti-symmetric under (\ref{tukor}), one of
them has no (lattice) momentum, in the other this quantity is
$\pi$. Above each vacuum there is a set of excited states which can
be considered as scattering states of dressed particles. 
The momenta of the
particles are quantized through a set of equations of
the BA type (\ref{hlbae1a}-\ref{hlbae1b}) 
(HLBAE).  The excitation energy is the sum of the energy
contributions of the individual particles and the momentum is the sum
of the contribution of the individual particles added to the momentum
of the corresponding vacuum.  The $S^z=0$ states are eigenstates of
the operation flipping all spins, and 
the eigenvalue of this operation is related to the chain length,
quantum number $I_0$ and the rapidities describing the excitations. 

%%%%%%%%%%%%%%%%%%%%%%%%%%%%%%%%%%%%%%%%%%%%%%%%%%%%%%%%%%%%%%%%%%%%
\section{The scaling limit}
\label{sec:skalalim}

\subsection{The relativistic spectrum}
\label{sec:spectr}

Before constructing the scaling limit we note, that certain quantities
(the momentum $Q$ and the reflection $\Sigma$) depend on the parity of $N$
(and $N/2$) but not on its magnitude. To make the procedure uniquely defined 
we carry out the $N\to\infty$ limit through $N$s being integer multiples of
four, in which case both in $Q$ and $\Sigma$ the $N$ can be replaced by zero,
and at the end of the section we discuss the consequences of other possible 
choices.

First, we show that (\ref{rellim}) defines the limit in which
the dispersion relation (\ref{ep}) describes relativistic particles. 
Note that the continuum momentum ($P$, $p$) is obtained from
the lattice momentum ($Q$, $q$) after division by the lattice spacing
$a$:
\be
\epsilon^2=\left({V\over a}{K\over\pi}\sinh\gamma\right)^2
\left(1-k^2\cos^2(ap)\right)
\approx 
\left({VKk\sinh\gamma\over\pi}\right)^2
\left(\left({k'\over ka}\right)^2+p^2\right)\,.
\ee
Here we expanded $\cos(ap)$ about 1 and now observe that in the
continuum limit ($a\rightarrow 0$) a relativistic dispersion relation
emerges, provided $\gamma\to0$ together with
\be
\left({VKk\sinh\gamma\over\pi}\right)=1\quad{\rm and}\quad 
\left({k'\over ka}\right)=M\,.
\ee
These requirements imply $V=2/\pi$ and (\ref{rellim}) (see Appendix
\ref{sec:lista}).  

It will prove to be useful to introduce the scaled rapidity variable,
$\vartheta$ which also makes the limiting procedure more transparent. 
It is defined in terms of the $\theta$ and $\gamma$ as (see also
(\ref{sigma0})):
\be\label{vartheta}
K+\vartheta \equiv
K\frac{2\theta}{\pi}\,,
\ee
while the energy $\varepsilon(\vartheta)=\epsilon(\theta(\vartheta))$
and momentum $p(\vartheta)={1\over a}q(\theta(\vartheta))$ can be
expressed as 
\be\label{sclr}
\varepsilon(\vartheta) = 
\frac{2K\sinh\gamma}{\pi^2}\frac{k'}{a}\frac{1}{\dn(\vartheta)} \\
\;\;;\hspace{.5in}
\frac{\sin(ap(\vartheta))}{a} =
\frac{k'}{a}\frac{\sn(\vartheta)}{\dn(\vartheta)}\,. 
\ee
In the scaling limit $\vartheta/K$ approaches 0 as $k\rightarrow 1$
which yields
\be\label{vt}
\pi\frac{\theta-\pi/2}{\gamma}\rightarrow\vartheta
\ee
\be
\varepsilon(\vartheta)\rightarrow\frac{k'}{a}\cosh(\vartheta) 
\;\;;\hspace{.5in}
p(\vartheta)\rightarrow\frac{k'}{a}\sinh(\vartheta), 
\ee
in accord with ${k'}/{a}\rightarrow M =$ constant. 

The total momentum 
also contains a macroscopic quantity (\ref{ptheta}) and can be written
in the continuum limit as,
\be\label{mom1}
P={Q(I_0)\over a}+\sum_h M\sinh(\vartheta_h)\,.
\ee
In view of (\ref{mom1}) the two sectors corresponding 
to $I_0=1/2$ and $I_0=0$ become
infinitely separated in momentum space lending a strong support to 
interpreting the $I_0=1/2$ resp.~$I_0=0$ excited states as excitations 
above the corresponding ($I_0=1/2$ resp.~$I_0=0$) vacua.
Technically while (\ref{mom1})
gives the true momentum for the $I_0=1/2$ sector in the  $a\to0$
limit, the continuum momentum of the $I_0=0$ sector needs to be
redefined as 
\be
\label{mom2}
P={Q(I_0)+\pi\over a}+\sum_h M\sinh(\vartheta_h)\,.
\ee
As an alternative, we might keep
both sets by redefining the lattice so that two sites form one
elementary cell. In this case the lattice momenta $0$ and $\pi$ are
equivalent, and the $Q(I_0)$ term can be dropped. In any case the 
energy and momentum measured relative to those of the corresponding vacuum are 
\be\label{relem}
E-E_0=\sum_h M\cosh(\vartheta_h)\,,
\quad\quad P=\sum_h M\sinh(\vartheta_h)\,.
\ee

\subsection{The vacua in the scaling limit}
\label{sec:vakumok}

The two lowest lying states are characterized purely by the density of
rapidities, $\sigma_0(\lambda)$ whose behavior in the scaling limit is
crucial for our arguments. The reason is that our considerations about
the excited states are based on the existence of a non-vanishing
background of rapidities (see also Appendix \ref{sec:strstr}). 
One can show that although $\sigma_0(\lambda)$ vanishes around $\pi/2$
(and diverges at $0$ and $\pi$), 
the density of the $\vartheta$'s is non-vanishing as required. 
The density $\varrho_0(\vartheta)$ of the rescaled
rapidities (\ref{vartheta}) can be expressed as
\be
N\sigma_0\left(\frac{\pi}{2K}(\vartheta+K)\right)d\lambda=
L\varrho_0(\vartheta)d\vartheta\,,
\ee
provided $d\vartheta=2Kd\lambda/\pi$ (here the l.h.s.~is the number
of $\lambda$'s in the interval $\{\lambda,\lambda+d\lambda\}$ at 
$\lambda=\pi/2+\pi\vartheta/2K$, and the r.h.s.~gives the number of 
$\vartheta$'s in the interval $\{\vartheta,\vartheta+d\vartheta\}$),
which leads to  
\be
\varrho_0(\vartheta)={1\over2\pi}M\cosh(\vartheta)\,.
\ee

The ({\ref{endifiscl}) difference in the energies of the two vacua is
obtained directly by evaluating the scaling limit of (\ref{endifi}).
We have to note, however, that (\ref{endifi}) is the leading term in a
large $N$ expansion, and as the expansion coefficients are functions
of $\gamma$ one has to see, if these neglected terms behave
properly. In Appendix \ref{sec:alapok} we show 
that (\ref{endifiscl}) is the leading term of a large $L$
expansion of the same quantity and the neglected terms decay faster.

Finally we note, that the reflection symmetry 
is not effected by the limiting process, i.e.~$\Sigma=\pm1$ in
the two states.

\subsection{Secular equations in the scaling limit}
\label{sec:seceq}

We define the secular equations of the limiting theory as the SL of the 
HLBAE of the lattice model. This limit exists 
but if it is meaningful is a delicate question: 
deriving the HLBAE two approximations has been used, 
integrating instead of summing over the rapidities, 
and neglecting the exponentially
small corrections to the close pairs. These have to be
justified even in the SL.
Replacing the sums with integrals require that the density of rapidities does
not vanish in the scaling limit (see the previous section) 
more over it can be shown that the error introduced remains negligible. 
Also the corrections to the close pairs remain exponentially small 
even in the SL as it is shown in Appendix \ref{sec:strstr}.

To derive the scaling limit of the HLBA we first observe, that the
energy and momentum diverge unless all $\theta$ scale to
$\pi/2$ as $\gamma\to0$ (i.e.~all $\vartheta$  are finite).
In this case we expect also some of the $\chi$'s to scale to $\pi/2$. 
For convenience we introduce their rescaled versions as
\be
\label{kappadef}
\kappa=\pi\frac{\chi-\pi/2}{\gamma}
\ee
It is straightforward to determine how the functions $\am$ , ${\cal F}$
and $\Phi$ behave in the scaling limit:
\bea
{1\over a}\left(\am(\vartheta+K)-{\pi\over2}\right)
&\stackrel{\gamma\rightarrow 0}{\longrightarrow}&
p(\vartheta)=M\sinh\vartheta\,,\\
{\cal F}\left({\Delta\vartheta\gamma\over\pi},\gamma\right)
&\stackrel{\gamma\rightarrow 0}{\longrightarrow}&
\phi\left({\Delta\vartheta\over\pi}\right)\,,\\
\Phi\left({\Delta\kappa\gamma\over\pi},{\gamma\over m}\right)
&\stackrel{\gamma\rightarrow 0}{\longrightarrow}&
\pi-2\tan^{-1}\left({\Delta\kappa\over\pi/m}\right)\,, \\
\Phi\left({\Delta\chi},{\gamma\over m}\right)
&\stackrel{\gamma\rightarrow 0}{\longrightarrow}&
0\hspace{1.5in}{\rm if}\;\;
\Delta\chi\to{\rm finite.}
\eea
After substituting the above expressions to the HLBA equations, their
scaling limit is obtained. The complex rapidities which remain at
finite distance from $\pi/2$ drop out from the equations, although they
do affect $S^z$ of the state as we will see in the next section. 
The equations for the holes and the complex parameters approaching
$\pi/2$ become 
\bea\label{geea}
Lp(\vartheta_h)&=&2\pi({\cal I}_h+I_0)-\sum_l^{n(\vartheta)}\phi\left(
{\vartheta_h-\vartheta_l\over\pi}\right)
+\sum_{\alpha}^{n(\kappa)}
2\tan^{-1}{\vartheta_h-\kappa_{\alpha}\over\pi/2}\,,\\
&&{\cal I}_h={1\over2}\left({n(\vartheta)-
2n(\kappa)\over2}\right)\modegy\,,  \\  
\label{geeb}
\sum_h^{n(\vartheta)}2\tan^{-1}{\kappa_{\alpha}-\vartheta_h\over\pi/2}
&=&2\pi{\cal J}_{\alpha}+\sum_{\beta}^{n(\kappa)}
2\tan^{-1}{\kappa_{\alpha}-\kappa_{\beta}\over\pi}\,,\\
&&{\cal J}_{\alpha}=\left({n(\kappa)-
n(\vartheta)+1\over2}\right)\modegy\,.
\eea
It is to be noted, that the quantum number $I_0$ directly appears in
the secular equations of the excited states.  As we have discussed
before, we interpret the two sets of solutions distinguished by the
value of $I_0$ as the excited states of the corresponding vacua.

\subsection{Multiplet structure of the states}
\label{sec:multipl}

A few simple solution of the above secular equations are given and the 
structure of solutions is discussed in Appendix \ref{sec:solstr}. Here
we discuss the degenerations in the spectrum developing in the SL.
As both the energy and the momentum depend on the rapidities of the holes only,
we expect degenerate multiplets corresponding to solutions differing
in their $\chi$ content only. As (\ref{geea}) and (\ref{geeb})
contain both the $\vartheta$ and $\kappa$ sets the solutions becoming 
degenerate can differ in the $\chi$'s not scaling to $\pi/2$
(and which disappeared from the equations).

In Appendix \ref{sec:solstr} we discuss the types of different solutions
of the HLBAE (\ref{hlbae1b}) and relate them to the solutions of 
(\ref{geeb}). We argue, that at a given number of holes, 
$n(\vartheta)$, each
having the form $\theta_h=\gamma\vartheta_h/\pi+\pi/2$,
to any number $0\leq n(\kappa)\leq n(\vartheta)/2$ (\ref{hlbae1b})
have $n(\vartheta)/2\smi n(\kappa)\spl1$ different solutions in which 
$n(\kappa)$
of the $\chi$s scale to $\pi/2$ like 
$\chi_{\alpha}=\gamma\kappa_{\alpha}/\pi+\pi/2$ with the 
common $\kappa_{\alpha}$ set being one solution of (\ref{geeb}). These
solutions differ in the number of $\chi$s not scaling to $\pi/2$.
It is clear than, that the number of 
different {\em solutions} of (\ref{hlbae1a}-\ref{hlbae1b})
becoming degenerated in the SL and given 
by one solution of the system (\ref{geea}-\ref{geeb}) is 
$n(\vartheta)/2\smi n(\kappa)\spl1$. 
The number of $\chi$s not scaling to $\pi/2$ can be
$0,1,\ldots,n(\vartheta)/2\smi n(\kappa)$, and the corresponding states
have spins $S^z=n(\vartheta)/2\smi n(\kappa),\ldots,1,0$ respectively.
Completing this set of states with the states
$S^z=-1,-2,\ldots,-(n(\vartheta)/2\smi n(\kappa))$ which are obtained by
flipping all spins we see, that the
degeneration of the {\em states} given by one solution of the equations
(\ref{geea}-\ref{geeb}) is $n(\vartheta)\smi 2n(\kappa)\spl1$ 
This corresponds exactly to SU(2) multiplets:
the states are characterized by two quantum number
$l=n(\vartheta)/2\smi n(\kappa)$ and $l\geq m\geq-l$, and the $m$ does not
influence the energy. This way we may consider the particles as `spin'
1/2 particles forming SU(2)  eigenstates with quantumnubers 
$l$ and $m$. (This picture will be refined when discussing the 
spin reversal symmetry.)  Apparently the
equations (\ref{geea}) and (\ref{geeb}) give the $l=m$ states directly
(as usual in SU(2) BA systems), and
to have the complete description of the $m<l$ states one should find
the $\chi$s not scaling to $\pi/2$, nevertheless these parameters do
not enter into the spectrum. This interpretation is consistent with
the expectation for the number of different solutions: the number of
nonequivalent solutions of (\ref{geeb}) given by (\ref{izoszam}) is
just the number of different $l=n(\vartheta)/2\smi n(\kappa)$, $l=m$
states of $n(\vartheta)$ spin 1/2 particles.
One has to note, however, that while the quantumnumber $m$ coincides
with the $S^z$ of the system, it can not be seen, if $S^2=l(l+1)$ or 
the quantity given by $l(l+1)$ is different from the square of the total
spin. Note also, that this SU(2) symmetry develops in the SL, but it is 
not present in the initial model.

\subsection{The two particle scattering matrix}
\label{sec:Smatrix}

The two particle scattering phase shifts of the physical particles can
be reconstructed from the secular equations. The method for this is
based on the idea, that the deviations of the particle momenta from
the free values can be interpreted as the phase shifts of the
particles scattering on each other \cite{AnLo2,Kor,DeLo1}. Consider a
two particle scattering-state on a ring. If the momenta are $p_1$ and
$p_2$ and the particles obey twisted boundary conditions with a twisting 
angle $\varphi$ ($\varphi=0$ for periodic and it is $\pi$ for the 
antiperiodic boundary condition), then the boundary condition requires  
$Lp_1+\delta_{12}=2\pi n_1+\varphi$ and 
$Lp_2-\delta_{12}=2\pi n_2+\varphi$ with $n_1$ and $n_2$ integers and
$\delta_{12}$ being the phase shift \cite{HaQuiBa,SuSha}. 
Writing the equations (\ref{geea}) in this form the phase shifts can be found.
A weakness of this procedure is that in most of the interesting cases 
$\varphi$ is not known 
(as the boundary conditions for the dressed particles can 
differ from those of the bare ones). The twisting could be seen directly 
in the equation describing one single particle. Since, however, the 
parity of the particle number is usually determined by that of the 
number of elements in the chain, the sectors with even and odd 
number of particles are disjoint in the sense, that a given system represents
either one or the other, and one can not see, if the twisting is the same 
in both sectors. (A trivial example 
for particle number dependent boundary conditions is given by a spin chain 
where the spin waves are described by Fermions through a Jordan-Wigner 
transformation.) Now we {\em suppose} periodic boundary conditions, 
i.e.\ $\varphi=0$, but we keep in mind, that our results hold up to a
rapidity independent constant (see also \cite{EssKo}).

A triplet state of two particles is described by two $\vartheta$s and
no $\kappa$s. For such a state Eq.\ (\ref{geea}) yields
\be\label{deltatr}
\delta^{tr}_{12}=\phi\left({\Delta\vartheta\over\pi}\right)+
\pi-2\pi I_0\,, \quad \Delta\vartheta=\vartheta_1-\vartheta_2\,,
\ee
where the $\pi$ comes from the parity prescription for the parameters
${\cal I}_h$.  A singlet state is characterized in addition to
$\vartheta_1$ and $\vartheta_2$ by a $\kappa$ for which Eq.\
(\ref{geeb}) yields $\kappa=(\vartheta_1+ \vartheta_2)/2$.  For this
case the above reasoning leads to
\be\label{deltasing}
\delta^{s}_{12}=\phi\left({\Delta\vartheta\over\pi}\right)-
2\tan^{-1}\left({\Delta\vartheta\over\pi}\right)-2\pi I_0\,.
\ee
From the phase shifts (\ref{deltatr}) and (\ref{deltasing})
the two particle $S$-matrix can be given as 
\be\label{smm}
\hat S(\Delta\vartheta)=-\exp\left\{i\left(2\pi I_0+
\phi\left({\Delta\vartheta\over\pi}\right)\right)\right\}
\left(\hat{P}_{tr}+{\Delta\vartheta+
  i\pi\over \Delta\vartheta-i\pi}\hat{P}_{s}\right)\,,
\ee
where $\hat{P}_{tr}$ and $\hat{P}_{s}$ are the 
projectors on the triplet resp.~singlet subspace of the two spins.
This $S$-matrix is consistent with
the SU(2) symmetry of the limiting model detected in the degenerations.

The scattering matrices (\ref{smm}) of the two sets of solutions 
differ in an overall sign due to the appearance of $I_0$.
To decide, if this difference is significant in the sense that it has 
physical consequences regarding the nature of the particles in the two 
sets of states, needs further considerations.

\subsection{Spin reversal symmetry}
\label{sec:reflsim}

Finally we have to discuss the question of the symmetry of the excited
states with respect to $\hat\Sigma$ of
(\ref{tukor}).  The two vacua are eigenstates with eigenvalues $\pm1$.
As we have mentioned it already, checking numerically a number of $S^z=0$
excited 
states of the finite lattice model (with finite $\gamma$) we have seen, that
these states are 
also eigenstates of $\hat\Sigma$
with the eigenvalue $\Sigma$ given by (\ref{sumall}-\ref{simsum}), 
but on that level we could not see an easy rule for
the eigenvalues. We have found some indications, however, that there
is a correlation between the spin-structure of the dressed
particles and the spin reversal symmetry. 
We have examined a number of solutions of different types
with two holes and one close pair of the $N=20$ lattice. It has been found
that in accordance with Appendix \ref{sec:solstr} 
the solutions form two classes: in class 1 the real part of the close
pair is in the vicinity of 
$(\theta_1+\theta_2)/2$ while in class 2 it is in the neighborhood of
$(\theta_1+\theta_2\pm\pi)/2$. In case of $I_0=1/2$ the solutions in class 1
are symmetric and the others, in class 2, are antisymmetric under
$\hat\Sigma$ while just the opposite is true for $I_0=0$
(Appendix~\ref{sec:numerics}). 

In the scaling limit the solutions of class 1
resp.~class 2 satisfy (\ref{sing}) resp.~(\ref{trip}) representing singlet
and triplet states. We believe that while increasing the value of $N$ the
symmetry properties do not change, and we conclude, that in this type 
of states the singlets have the symmetry of the vacuum, and the symmetry
of the triplets is the opposite. Accepting the validity of (\ref{sumall}) 
(and so that of (\ref{simsum}))
we can have $\Sigma$ also for the $n(\theta)=4$, $n(\chi)=2$ states.
Taking the results of Appendix \ref{sec:solstr} we find that
\be\label{genszigma}
\Sigma=\Sigma(I_0,N,l)=(-1)^{(2I_0-1)+N/2-l}=\Sigma(I_0,N)(-1)^l
\ee
with $l=0,\,1,\,2,$ corresponding to singlet, triplet and quintuplet
configurations, respectively. We think, this is an indication, that
(\ref{genszigma}) is generally true.  

If we perceive the reflection symmetry
as the product of the symmetry of the vacuum and that of the spin structure
of the excitations, the latter should be 
\be\label{spart}
\Sigma=(-1)^l
\ee
i.e.\ we have to consider now the singlets to be symmetric and the triplets 
anti-symmetric. 
This is just the opposite as in the case of two ordinary 
SU(2) spins, and resembles 
a $q$-deformed SU(2) structure at $q=-1$ 
\cite{Balog}. It can be shown in general \cite{Wo}, that     
defining the spin operators for a system of particles as
\be
\sigma^{\pm}_2=\sum(-1)^{(j-1)}\sigma^{\pm}_j\quad\quad 
\sigma^z_2=\sum\sigma^z_j
\ee
(with $\sigma_j$ being the Pauli matrices acting on the spin labeled by $j$)
results an SU(2) structure (equivalent to a $q$-deformed SU(2) at $q=-1$)
in which the $S^z=0$ members of the multiplets have the eigenvalue 
(\ref{spart}). Particularly in the case of two spins
the $\sigma_2^z=0$
member of the triplet is $\frac{1}{\sqrt2}(|\uparrow\downarrow\rangle-
|\downarrow\uparrow\rangle)$, while the singlet is
$\frac{1}{\sqrt2}(|\uparrow\downarrow\rangle+
|\downarrow\uparrow\rangle)$. 

We have to note, that the symmetry properties of the limiting theory are in 
strong analogy with that of an ordinary XXX Heisenberg chain. The 
eigenstates of this model are $S^2$ and $S^z$ eigenstates with eigenvalues
$l(l+1)$ and $m$, respectively, with $l\geq|m|$ integers if $N$ is even.
It is known, that the $m=0$ states of an even number of spins
are eigenstates of $\hat\Sigma$ with an eigenvalue analogous to 
(\ref{genszigma})
\be\label{genszigmap}
\Sigma_{XXX}=(-1)^{N/2-l}\,.
\ee
This
is a consequence of the fact, that any such state can be given as linear
combination of states built up as products of $l$ triplet and $N/2-l$
singlet pairs \cite{SuFa}. The ground state (the vacuum) is singlet, thus
all the singlet excitations have the same symmetry as the vacuum
and the triplets have the opposite. 
Thus we may consider the degenerated states as forming normal SU(2) multiplets
of the $N$ spin of a the chain, but equally well we may think of these 
states as the above described multiplets of the
{\em dressed particles}, and all this is a direct
consequence of the SU(2) symmetry of the XXX Hamiltonian. In the case of our 
limiting theory  
there is no Hamiltonian, and the symmetry
is recognized in an indirect way: the 
SL is taken through states not showing the SU(2) structure, 
but this symmetry appears, as degenerations characteristic for 
this symmetry develop. Now we see, that the reflection symmetry
properties complete this structure. 

\subsection{Boundary conditions in the limiting theory}
\label{sec:boundary}

In the previous part of the Section we supposed, that  the $N\to\infty$
limit is taken through integer multiples of four ($N=4\cal N$). 
Actually in order to be able to define the limit of the momentum uniquely and
to obtain the parameters in the limiting equations together with the
properties of the excited states properly defined, we have to fix the
way of the $N\to\infty$ limit, but we have four different possibilities:
$N=4\cal N+\nu$, $\cal N\to\infty$ and $\nu=0,1,2,3$. In these cases
(\ref{pi0}) and (\ref{genszigma}) are
\bea
\label{pi02}
Q(I_0,N)&\equiv&Q(I_0,\nu)=2\pi(I_0-1/2+\nu/4)\,, \\
\label{tukorertek2}
\Sigma(I_0,N,l)&=&\Sigma(I_0,\nu,l)=(-1)^{(2I_0-1)+\nu/2+l}
\eea
(the latter applying for $\nu=0,2$ only, as $\hat\Sigma$ has no eigenstates 
among the BA states of the $N$=odd chain).
Now we want to investigate the $\nu=1,2,3$ cases. 

First consider the $\nu=2$ case. Now we have to define the momentum in the 
continuum limit as
\be\label{mom3}
P={Q(I_0,2)-\pi\over a}+\sum_h M\sinh(\vartheta_h)
\ee
or
\be\label{mom4}
P={Q(I_0,2)\over a}+\sum_h M\sinh(\vartheta_h)
\ee
to pick the $I_0=1/2$ or $I_0=0$ states respectively. In this procedure
all the results are exactly the same as before, except those concerning 
the reflection symmetry (and as we shall see the $S$-matrices must be affected 
too). Now, according to (\ref{tukorertek2}) ((\ref{tukorertek})) the secular 
equations (\ref{geea}) with $I_0=0$ describe the excitations of the 
reflection symmetric vacuum, and $I_0=1/2$ gives the excitations of the 
antisymmetric one (just the opposite as for $\nu=0$). 
The nature of the excitations should be given 
by the symmetry of the corresponding vacuum, and the secular equations
are the quantizations of the momenta of the individual particles. Now the
fact, that the same set of particles (say a singlet pair above one of the 
vacua) may obey two different quantization condition, can be interpreted so
that in the two cases the {\em boundary conditions} are different:
if they are
periodic in one case, they are anti-periodic in the other
(the difference in the twisting is $\pi$).
The two 
particle scattering matrices given by (\ref{smm}) are obtained supposing
periodic boundary conditions. From the very same secular equations but 
supposing anti-periodic boundary conditions the same $S$-matrices multiplied
by $(-1)$ are obtained, i.e.~interpreting the limiting theories 
for $\nu=0$, $I_0=0(1/2)$ and $\nu=2$, $I_0=1/2(0)$ as being the same ones
but with different boundary conditions, is consistent with the $S$-matrices
obtained. It is also consistent with the reflection symmetry of the $S^z=0$
states: we checked numerically (for $N=18$), that also in the $\nu=2$ case
the two particle states scaling to singlets have the same symmetry as the 
vacuum. 

Now it is not hard to see, that with slight modifications of the prescription
of the limiting process for the momentum we can consistently define the 
scaling limit also for $\nu=1$ and $\nu=3$. This way, together with the two 
possible values of $I_0$ we can obtain four cases. Also these are described
by the equations (\ref{geea}-\ref{geeb}) but with the restriction, that
$n(\vartheta)$ must be odd. Also in these cases the eigenstates form 
SU(2) multiplets 
corresponding to the different combinations of $n(\vartheta)$ spins of 
length $1/2$ and
Eqs.~(\ref{geea}-\ref{geeb}), just as in the
previously discussed cases, give directly the 
$l=m=n(\vartheta)/2-n(\kappa)$ states.
We interpret these four sets of states as the excited states with odd
numbers of particles above the two vacua, 
both with two different boundary conditions. Unfortunately, as none 
of these sets contain the corresponding vacuum, we can not tell, 
which values of the parameters $I_0$ and $\nu$ correspond to which vacuum.
Now, however, the boundary conditions can be seen directly: the equations of 
the one particle states read
\be
Lp(\vartheta)=2\pi(I_0+1/4+{\rm integer})
\ee
indicating, that these particles obey twisted boundary conditions with
twisting angle $2\pi(I_0+1/4)$ \cite{HaQuiBa,SuSha}.

A possible explanation of the fact that for different $\nu$s we get 
different limits
can be the following. In many cases in order to define the local
operators in a naive continuum limit one has to group the lattice points 
into elementary cells of certain length \cite{WoEcTr,WoFo1,WoFo2},
which will be the points of the continuum.
If the chain length is not an integer multiple of the size of the
elementary cell, necessarily boundary terms occur in the limit
which can cause a rapidity independent shift in the phase (twisting) as 
a particle is taken around the ring.
Now it seems, that the size of the elementary cell is four, and so there
are four different possibilities as far as the `surface terms' are concerned.
In addition to this the number of particles must have the same parity as the 
chain-length (the vacua are spinles, while the particles are spin 1/2 particles).
These together add up to give the variety of sectors discussed above.

\section{Acknowledgment}

We are grateful to J.~Balog, P.~Forg\'acs, L.~Palla and J.~S\'olyom
for illuminating discussions. T.H.~was supported in part by the
U.S.\ Department of Energy under contract \#DE-FC02-94ER40818 and by
the Hungarian National Science Fund OTKA, grant No.~T19917. F.W.\ was
supported by the Hungarian National Science Fund OTKA, grants
No.~T014443 and T022607.

%*%*%*%*%*%*%*%*%
\appendix
\section{}
\label{sec:lista}

In this appendix we list those definitions and properties of the
elliptic integrals and functions \cite{GR} we use in the bulk of the
paper.

\bea
F(\phi,k)&\equiv&\int_0^\phi\frac{d\alpha}{\sqrt{1-k^2\sin^2\alpha}}
\\
K&\equiv&\int_0^{\frac{\pi}{2}}(1-k^2\sin^2\alpha)^{-\frac{1}{2}}d\alpha
\\
K'&\equiv&\int_0^{\frac{\pi}{2}}(1-k'^2\sin^2\alpha)^{-\frac{1}{2}}d\alpha
\\ k^2+k'^2 &=& 1\\ k_1 &=& \frac{(1-k^{\prime})^2}{k^2} \\ K' &=&
\frac{\pi}{2}(1+\frac{1}{4}k'^2+\ordo{k'^4}) \\ K
&=&\log\frac{4}{k'}+\frac{1}{4}(\log\frac{4}{k'}-1)k'^2+\ordo{k'^{4-0}}
\\
\frac{K'}{K}&=&\frac{\pi}{2\log\frac{4}{k'}}(1+\frac{k'^2}{4\log\frac{4}{k'}}
+ \ordo{k'^{4-0}} ) \\ \gamma &=& \pi\frac{K'}{K} \\ k' &=&
4e^{-\frac{\pi^2}{2\gamma}}-16e^{\frac{3\pi^2}{2\gamma}}+\ldots \\ K
&=& \frac{\pi^2}{2\gamma}+\frac{2\pi^2}{\gamma}
e^{-\frac{\pi^2}{\gamma}}+\ldots \\ K' &=&
\frac{\pi}{2}(1+4e^{-\frac{\pi^2}{\gamma}}+\ldots) \\
{\partial\,\am(u,k)\over\partial\,u}&=&\dn(u,k)\\
{\partial\,\sn(u,k)\over\partial\,u}&=&\cn(u,k)\dn(u,k)\\
{\partial\,\cn(u,k)\over\partial\,u}&=&-\sn(u,k)\dn(u,k)\\
{\partial\,\dn(u,k)\over\partial\,u}&=&-k^2\sn(u,k)\cn(u,k)\\ 
\sn(u,k)&=& \sin\am(u,k) \\ 
\dn^2(u,k) &=& 1-k^2\sn^2(u,k) \\ 
\dn(\pm K+u,k)&=& k'\frac{1}{\dn(u,k)} \\ 
\cn(\pm K+u,k)&=&\mp k'\frac{\sn(u,k)}{\dn(u,k)} \\ 
F(\phi,k)&\stackrel{k\rightarrow1}{\longrightarrow}& \log\tan\left(\frac{\phi}{2}+\frac{\pi}{4}\right)\\ 
\sn(u)&\stackrel{k\rightarrow 1}{\longrightarrow}& \tanh(u) \\
\dn(u)&\stackrel{k\rightarrow 1}{\longrightarrow}& \frac{1}{\cosh(u)}
\eea

\section{}
\label{sec:alapok}

Here we analyze the behavior of the vacuum rapidity sets in the
SL and give a derivation of the energy difference between
the two vacua.
 
First one can see, that if $\gamma\to0$ at fixed $N$ the number of
rapidities is fixed and they condense at the ends of the interval
$\{0,\pi\}$, (except the $\lambda=\pi/2$ element in one of the sets),
and the interior of the interval becomes empty. In this limit if the
rapidities are rescaled $x=\lambda/\gamma$ (if $\lambda<\pi/2$) and
$x=(\lambda-\pi)/\gamma$ (if $\lambda>\pi/2$), the rapidity sets of
the singlet ground-state resp.~the lowest energy triplet
state of the XXX chain are recovered.  If however, the
$\gamma\to0$ limit is performed under the condition
(\ref{rellim'}), although most of the rapidities condense at $0$ and
$\pi$ but the interior of the interval will not become empty, and the
properly scaled rapidities have a finite density at $\pi/2$, as
discussed in \ref{sec:vakumok}. This is an important condition for the
applicability of the method deviced by \cite{deVW} to calculate the
energy difference of the two vacua.

The expression for the energy difference of the two vacua given in
\cite{deVW} gives the leading terms of a double series of the powers
of $1/\sqrt{N}$ and $k_1^{N/2}$ in which the coefficients are
functions of $\gamma$. If $\gamma\to0$ at fixed $N$ this expansion
breaks down as $k_1\to1$ and also the expansion coefficients may
explode. (While the expansion breaks down, the true energy difference
will be $\propto1/N$.) For this reason one has to be very careful, and
if our (\ref{rellim}-\ref{rellim'}) limit is performed, the limit
is to be calculated in all terms. Evaluating the terms presented in
\cite{deVW} indicates that the limiting energy difference is indeed a
function of $L$ (just as (\ref{endifiscl}) is). In the following we
present a simple derivation of (\ref{endifiscl}) in which one can see,
that the neglected terms decay faster.

According to \cite{deVW} the energy difference is given by
\be\label{kulonbsegek} \Delta E\simeq -4{\rm
Re}\int\limits_{0+i\gamma/2}^{\pi+i\gamma/2}
\epsilon(\theta)\sigma_0(\theta)e^{iNq(\theta)}d\theta 
\ee 
where
$\epsilon(\theta)$, $\sigma_0(\theta)$ and $q(\theta)$ are given by
(\ref{epstheta}), (\ref{sigma0}) and (\ref{ptheta}) respectively, and
the neglected terms are of the type 
\be\label{elhagyva}
\int\limits_{0+i\gamma/2}^{\pi+i\gamma/2}
f(\theta)e^{iNnq(\theta)}d\theta \quad\quad(n\geq2\ {\rm integer})\,.
\ee 
The function $q(\theta+i\gamma/2)$ has an extremum at
$\theta=\pi/2$, and we can use the saddle-point method. Introducing
the variable $\vartheta$ of (\ref{vartheta}) we find \cite{AbrSte},
that around $\theta=\pi/2$ 
\be 
q(\theta+i\gamma/2)={1\over
i}\ln\left({1-k'\over k}\right)+ {ik'\over2}\vartheta^2\,.  
\ee
i.e.~(as $k'/a\to M$): 
\be
iNq(\theta+i\gamma/2)=N\,\ln\left({1-k'\over k}\right)-
{LM\over2}\vartheta^2\,.  
\ee 
It is clear, that for large $L$ to pick
up the leading contribution we may use this expression in the
integral, and we may expand also the $\epsilon$ and $\sigma$ in terms
of $\vartheta$.  This leads to 
\be 
\Delta E\simeq
{8NK\sh\gamma(k')^2\over a\pi^3} \left({1-k'\over k}\right)^N
\int\limits_0^{\infty}\vartheta^2e^{-LM\vartheta^2/2}d\vartheta 
\ee
which yields (\ref{endifiscl}) indeed. We can also see, that the
neglected terms (\ref{elhagyva}) are 
\be 
\propto e^{-nLM} 
\ee 
i.e.~for
large enough size $L$ the energy difference decays as
(\ref{endifiscl}).

\section{}
\label{sec:strstr}

In the bulk of the paper we derived the secular equations of the limiting model as
the limit of the secular equations of the lattice model. This
procedure is meaningful only if the approximations used to derive the
original equations remain valid in the SL too.

One of the approximations applied is the replacement of summations by
integrals. The error introduced this way can be calculated by the
method of \cite{deVW}, and as it leads to the evaluation of formulae
of the type (\ref{kulonbsegek}), we may conclude, that this type of
corrections are exponentially small in $LM$.

Another subtle point of the derivation of the higher level BA
equations is the representation of each close pair by a single number,
namely by its center. As it is known, the close pairs of the original
rapidity set are defined by those $\chi$ solutions of the higher level
BA equations, for which ${\rm Im}\chi<\gamma/2$ through the rule
(\ref{kozeli}). This approximation is allowable, if $\exp\{-N\Omega\}$
will remain small in the SL. Now we show, that this is
indeed the case.  Actually according to \cite{BVV} 
\be 
\ln\left\vert
z^+-z^--i\gamma\right\vert\simeq-N\Omega(z^+,\gamma)\,, 
\ee 
with 
\be
\Omega(z^+,\gamma)=\frac{1}{2} \sum_{m=-\infty}^\infty\ln
\left\{\frac{\cosh\frac{\pi}{\gamma}(x+m\pi)+
\sin(\frac{\pi}{\gamma}y)}{\cosh\frac{\pi}{\gamma}(x+m\pi)-
\sin(\frac{\pi}{\gamma}y)}\right\}\,,\quad(z^+=x+iy,\ \ \
y\leq\gamma)\,.  
\ee 
In studying the behavior of the $\Omega$ one has
to distinguish three cases (see Appendix \ref{sec:solstr}):
\begin{itemize}
\item[1.] The $x$ scales to $0$ or $\pi$: 
\be
x=\cases{\xi\gamma/\pi&$\xi\geq0$\cr \pi+\xi\gamma/\pi&$\xi<0$\cr}
\ee 
In this case, as also $y\propto\gamma$, $\Omega\to\,$finite, and
$N\Omega\to\infty$, i.e.~$z^+\!-\!z^-\!-\!i\gamma\to0$ exponentially
fast in $N$.
\item[2.] Another possibility is, that $x$ is finite but
$x\not\to\pi/2$. It is not hard to see, that in this case 
\be
N\Omega\simeq\frac{LM}{2}e^{\frac{\pi}{\gamma}\vert\frac{\pi}{2}-x\vert}
\sin(\frac{\pi}{\gamma}y)\,.  
\ee 
Also this expression diverges as
$\gamma\to0$.
\item[3.] Finally, if $x$ scales to $\pi/2$, i.e.~ $x=\pi/2+\gamma{\rm
Re}\kappa/\pi$ (and $y=\gamma{\rm Im}\kappa/\pi+\gamma/2$ with ${\rm
Im}\kappa<\pi/2$), one finds, that 
\be 
N\Omega=LM\cosh({\rm
Re}\kappa)\cos({\rm Im}\kappa)\,.  
\ee
\end{itemize}

\noindent The close pairs described as case 1.~and 2.~form exact
2-strings or quartets in the scaling limit, but they disappear from
the secular equations. Nevertheless they play an important role in
determining the multiplet structure of the states. The close pairs of
case 3.~are in strong analogy with the close pairs of a finite lattice
model in the sense, that they form exact 2-strings or quartets in the
infinite size limit only. As, however, the deviation is exponentially
small in the size of the system, the representation of the close pairs
by their centers is justified.

\section{}
\label{sec:solstr}

In this Appendix we discus the structure of the solutions of
(\ref{hlbae1b}), what we write in the form
\be\label{anizo}
\prod_{h=1}^{n(\theta)}{\sin(\chi_j-\theta_h+i\gamma/2)\over
\sin(\chi_j-\theta_h-i\gamma/2)}=-\prod_{k=1}^{n(\chi)}
{\sin(\chi_j-\chi_{k}+i\gamma)\over \sin(\chi_j-\chi_{k}-i\gamma)}\,. 
\ee
We give an account of their different types, and relate their 
$\gamma\to0$ limit to the solutions
of (\ref{geeb})
\be\label{izo}
\prod_{h=1}^{n(\vartheta)}{\kappa_j-\vartheta_h+i\pi/2\over
\kappa_j-\vartheta_h-i\pi/2}=-\prod_{k=1}^{n(\kappa)}
{\kappa_j-\kappa_{k}+i\pi\over \kappa_j-\kappa_{k}-i\pi}\,. 
\ee
In this
analysis we shall keep the scaled rapidities $\vartheta_h$ fixed 
while seeking for solutions to $\chi_j$ as a perturbative expression in $\gamma$:
\be 
\theta_h=\frac{\pi}{2}+\frac{\gamma}{\pi}\vartheta_h\,, 
\;\;\;\;\;\;\;
\chi_j = \chi_{j0}+\frac{\gamma}{\pi}x_j + {\cal O}(\gamma^2).
\ee
Note that $\chi_{j0}=\frac{\pi}{2}$ and $x_j=\kappa_j$ for the
rapidities which scale to $\pi/2$ as was defined in (\ref{kappadef}).

We label the types of solutions by three numbers
$\{H,l,m\}$.  In this label the first number is $H=n(\vartheta)$, the
number of particles, the last one is $m=S^z=n(\vartheta)/2-n(\chi)$,
i.e.~the spin of the corresponding state, and the middle one is given
by $l=n(\vartheta)/2-n(\kappa)$. This quantum number is specified in
the bulk of the paper, when the multiplet structure is discussed.
(Note, that $l-m=n(\chi)-n(\kappa)\geq0$ is the number of $\chi$s 
{\em not} scaling
to $\pi/2$.)

We analyze in detail the states with $n(\vartheta)$ being equal to two
and four.

\bfl 
{\it States with $n(\vartheta)=2$} 

{$n(\chi)=0$}
\efl
The simplest such state is the one, in which there is no $\chi$. The
label of this state is $\{2,1,1\}$.

\bfl
{$n(\chi)=1$}
\efl
The equation 
\be 
\prod_{h=1,2}{\sin(\chi-\theta_h+i\gamma/2)\over
\sin(\chi-\theta_h-i\gamma/2)}=1 
\ee 
has two solutions: 
\bea
\chi&=&\frac{\theta_1+\theta_2}{2}\,,\label{sing}\\
\chi&=&\frac{\theta_1+\theta_2}{2}-\frac{\pi}{2}\modpi\,.\label{trip}
\eea 
In the first $\chi$ scales to $\pi/2$: 
\be\label{kikappa}
\chi=\frac{\gamma\kappa}{\pi}+\frac{\pi}{2} 
\ee 
with 
\be
\kappa={\vartheta_1+\vartheta_2\over2}\,, 
\ee 
in the second
not. According to our notations the two solutions are labeled by
$\{2,0,0\}$ and $\{2,1,0\}$, respectively.

Note, that in the SL the $\{2,1,1\}$ and $\{2,1,0\}$ solutions 
pairwise become degenerate, as the equations fixing the $\vartheta_h$'s
will be exactly the same for them.
 
\bfl 
\leftline{\it States with $n(\vartheta)=4$} 

{$n(\chi)=0$}
\efl
The simplest state is the one with no $\chi$. This is labeled by
$\{4,2,2\}$.

\bfl
{$n(\chi)=1$}
\efl
The states with one $\chi$ are characterized by the four nonequivalent
solutions of the equation 
\be
\prod_{h=1}^4{\sin(\chi-\theta_h+i\gamma/2)\over
\sin(\chi-\theta_h-i\gamma/2)}=1\,.  
\ee 
Three of these are well
approximated (in the $\gamma\to0$ limit are exactly given) by the
three (real and finite) solutions of the equation 
\be\label{egykappa}
\prod_{h=1}^{4}{\chi-\theta_h+i\gamma/2\over
\chi-\theta_h-i\gamma/2}=1\,,\quad\quad \left(
\prod_{h=1}^{4}{\kappa-\vartheta_h+i\pi/2)\over
\kappa-\vartheta_h-i\pi/2)}=1\,, \right) 
\ee 
and one $\chi$ is given in
leading order by 
\be
\chi=\frac{1}{4}\sum_{h=1}^4\theta_h-\frac{\pi}{2}\modpi=
\frac{\gamma}{4\pi}\sum_{h=1}^4\vartheta_h\modpi\,.  
\ee 
The
first three solutions have the label $\{4,1,1\}$, while the fourth is
$\{4,2,1\}$.

\bfl
{$n(\chi)=2$}
\efl
The solutions for the equations of two $\chi$s 
\be
\prod_{h=1}^4{\sin(\chi_j-\theta_h+i\gamma/2)\over
\sin(\chi_j-\theta_h-i\gamma/2)}=-\prod_{k=1}^2
{\sin(\chi_j-\chi_{k}+i\gamma)\over \sin(\chi_j-\chi_{k}-i\gamma)} 
\ee
are of three type. The set of first type coincides in the $\gamma\to0$
limit with the set of solutions of the equations 
\be
\prod_{h=1}^4{\chi_j-\theta_h+i\gamma/2\over
\chi_j-\theta_h-i\gamma/2}=-\prod_{k=1}^2
{\chi_j-\chi_{k}+i\gamma\over \chi_j-\chi_{k}-i\gamma}\,, 
\ee
i.e.~they are given through (\ref{kikappa}) by the solutions of 
\be
\prod_{h=1}^4{\kappa_j-\vartheta_h+i\pi/2\over
\kappa_j-\vartheta_h-i\pi/2}=-\prod_{k=1}^2 {\kappa_j-\kappa_{k}+i\pi\over
\kappa_j-\kappa_{k}-i\pi}\,.  
\ee 
These solutions are labeled by
$\{4,0,0\}$. (These solutions can be given even analytically. The two
nonsingular solutions (i.e.~in which the roots are not at the singularities 
of the equations) are of the form
\be
\kappa_{1,2}=\frac{1}{4}\sum_h^4\vartheta_h\pm\Delta\vartheta\,,
\ee
with $(\Delta\vartheta)^2$ given by 
\be
\left((\Delta\vartheta)^2\right)_{1,2}=\frac{1}{6}\left(-{\cal B}\pm\sqrt{
{\cal B}^2+12{\cal C}}\right)
\ee
with
\be
{\cal B}=2\left(\frac{\pi}{2}\right)^2+
\sum_{h<h'}(\delta\vartheta_{h})(\delta\vartheta_{h'})\,,
\quad
{\cal C}=\left(\frac{\pi}{2}\right)^4-2\left(\frac{\pi}{2}\right)^2
\sum_{h<h'}(\delta\vartheta_{h})(\delta\vartheta_{h'})+
\prod_h(\delta\vartheta_{h})
\ee
and
\be
\delta\vartheta_{h}=
\vartheta_h-\frac{1}{4}\sum_{h'}(\delta\vartheta_{h'})\,.
\ee
(The singular solutions are non-physical ones as they lead to vanishing
wave-function.)) 

Another set of solutions is given by 
\bea
\chi_1&=&\frac{\gamma\kappa}{\pi}+\frac{\pi}{2}\,,\nn\\
\chi_2&=&\frac{1}{2}\left(\sum_{h=1}^4\theta_h-2\chi_1\right)-
\frac{\pi}{2}\modpi
\frac{\gamma}{2\pi}\left(\sum_{h=1}^4\vartheta_h-2\kappa\right)\modpi\,, 
\eea
with $\kappa$ being one of the solutions
of (\ref{egykappa}).  The label of these three solutions is
$\{4,1,0\}$. Finally there is one solution labeled by $\{4,2,0\}$, for
which 
\be
\chi_{1,2}=
\frac{1}{4}\sum_{h}\theta_{h}
\pm i\tanh^{-1}\left(1/\sqrt{3}\right)+\ordo{\gamma}\,.  
\ee

The $\{4,2,2\}$, $\{4,2,1\}$ and $\{4,2,0\}$ solutions become degenerate
in the SL as the equations determining their $\vartheta_h$'s become identical.
For the same reason the $\{4,1,1\}$ and $\{4,1,0\}$ solutions form degenerate
pairs having the same $\kappa$.

\bigskip
\leftline{\it The general case}

Now we attempt to classify the solutions of the equations (\ref{anizo})
in the ($n(\theta)\geq2n(\chi)$) general case. It is clear, that in
the scaling limit in each solution some of the $\chi$s (say
$n(\kappa)\leq n(\chi)$ of them) will scale to $\pi/2$ in the form
(\ref{kikappa}) with the $\kappa$s satisfying (\ref{izo}) 
(with $n(\vartheta)=n(\theta)$).  Now our task is to give the number of
solutions at given $n(\theta)$, $n(\chi)$ and $n(\kappa)$, i.e.~the
number of solutions labeled by the same label $\{H,l,m\}$.  Giving
this number we have to make two points. First we suppose, that an
equation (\ref{anizo}) has 
\be
N_{XXZ}(n(\theta),n(\chi))={n(\theta)\choose n(\chi)} 
\ee
nonequivalent solutions (one solution being a set of $\chi$s).  This
statement is strongly related to the question if the BA yields a
complete set of eigenstates for the easy axis XXZ Heisenberg chain.
Our other assumption is, that an equation of the type (\ref{izo}) has
\be\label{izoszam} 
N_{XXX}(n(\vartheta),n(\kappa))=
{n(\vartheta)\choose n(\kappa)}-{n(\vartheta)\choose n(\kappa)-1} 
\ee
different solutions. This is related to the question of completeness
of the BA solutions for the isotropic case (in which a subset of the
HLBAE are again of the type (\ref{izo}). The difference of the two
statements originates from the fact, that in the isotropic case the BA
gives the $S^2=S^z(S^z+1)$ states only (i.e.~not all of the states),
while in the anisotropic case $S^2$ is not a good quantum number, and
the BA should give all the $S^z\geq0$ states. Now our conjecture is,
that the number of those solution of (\ref{anizo}) in which
$n(\kappa)(\leq n(\chi))$ of the $\chi$s scale to $\pi/2$ in the
scaling limit is again (\ref{izoszam}) (with
$n(\theta)=n(\vartheta)$), and in each solution the $n(\kappa)$ of the
$\chi$s which scale to $\pi/2$ are given by the different sets of
$\kappa$s solving (\ref{izo}). This means, that if $n(\chi)(\geq
n(\kappa))$ and a $\kappa$ set solving (\ref{izo}) are given, there is
no further freedom, and there is only one possibility to complete the
set of $\chi$s generated by the $\kappa$s to a set solving
(\ref{anizo}).  In other words, in a class of solutions $\{H,l,m\}$
there are $N_{XXX}(H,\frac{1}{2}H-l)$ different states uniquely
determined by the different solutions of (\ref{izo}) independently of
the value of m, (for which the only requirement is $l\geq m\geq0$).
This can be seen in the cases discussed, and is supported by the fact,
that 
\be 
\sum_{n(\kappa)=0}^{n(\chi)}N_{XXX}(n(\theta),n(\kappa))
=N_{XXZ}(n(\theta),n(\chi))\,.  \ee

A consequence of this is that in the scaling limit the solutions 
can be collected into groups of degenerate states: at a given $H$ and $l$
the members of such a group 
are characterized by common $\vartheta_h$ and $\kappa_j$ sets but 
differ in $m$ what runs from 0 to $l$.

\section{}
\label{sec:numerics}

In this appendix we show some solutions of the BAE of the finite lattice which
has $S^z=0$ and we give the symmetries of these states under $\hat\Sigma$.  
In these examples the length of the chain is always 20, the number of holes is
2, and we have one close pair. 
The equation (\ref{BAe}) was solved by iteration which was started from the
vicinity of the solutions of the HLBAE (\ref{hlbae1a}-\ref{hlbae1b}).
After this the amplitudes (\ref{BAstate}) of some configurations (which we
think dominate) were calculated and it was found that any of these amplitudes
was $\pm 1$ times the amplitude of the corresponding reflected configuration.
(The relative error was $10^{-6}$ but it is plausible that the 
reflection symmetry
holds exactly and the error comes only from the numerics.) 
The iteration converges fast also for much larger $N$, but the time needed for
the calculation of the amplitudes grows rapidly with $N$ resulting a strong
upper limit for the number of lattice sites we can handle. 
For illustration, Table~\ref{solutions} shows some of the solutions we 
have found.
There are two types of solutions: in type 1 the real part of the close pair
$v_c$ is between the two holes, while in type 2 it is located outside.
It can be seen in Table~\ref{solutions} that in case of $I_0={1/2}$
type 1 states are 
symmetric and type 2 ones are antisymmetric under $\hat\Sigma$ while the
opposite is true for $I_0=0$.
It should be noted that the connection between the type, $I_0$ and $\Sigma$
is independent of the interval the rapidities are collected in because
shifting one of the $\theta$s by $\pi$ both the position of $\chi$ compared
to the holes and the value of $I_0$ changes \cite{ViWo}. 
In our choice ($0<v<\pi$) type 1 resp.~type 2 solutions scale to singlet
(\ref{sing}) resp.~triplet (\ref{trip}) states as it is indicated in
Table~\ref{solutions}.  

\begin{table}
{\footnotesize
\begin{tabular}{cc}
\begin{tabular}{|c|c|c|} \hline
  $I$   & $v,~\theta$   & $\gamma,~v_c,~I_0,~\Sigma$ \\ \hline \hline
% data3  
   -0.5 &    0.0462991  &                                       \\
   -1.5 &    0.1481864  & $\gamma=1$                            \\
\bf-2.5 & \bf0.2663213  &                                       \\
   -3.5 &    0.4241941  & $v_c= 1.5046086 \pm i0.6433822$       \\
   -4.5 &    0.7049447  & ($\rightarrow$ singlet)               \\
   -5.5 &    2.4271298  &                                       \\
\bf-6.5 & \bf2.7097631  & $I_0={1 \over 2}$                     \\
   -7.5 &    2.8692656  &                                       \\
   -8.5 &    2.9880692  & $\Sigma=+1$                           \\
   -9.5 &    3.0904861  &                                       \\ \hline
% data15
   -0.5 &    0.1361835  &                                       \\
   -1.5 &    0.3022448  & $\gamma=1.5$                          \\
   -2.5 &    0.5124295  &                                       \\
   -3.5 &    0.8528326  & $v_c= 0.8744719 \pm i0.7496363$       \\
   -4.5 &    1.6684991  & ($\rightarrow$ triplet)               \\
\bf-5.5 & \bf2.2976980  &                                       \\
\bf-6.5 & \bf2.6042345  & $I_0={1 \over 2}$                     \\
   -7.5 &    2.8110406  &                                       \\
   -8.5 &    2.9773204  & $\Sigma=-1$                           \\
   -9.5 &    3.1276728  &                                       \\ \hline
\end{tabular}
&
\begin{tabular}{|c|c|c|} \hline
  $I$   & $v,~\theta$   & $\gamma,~v_c,~I_0,~\Sigma$ \\ \hline \hline
%data14
   -1.5 &    0.1438185  &                                       \\
   -2.5 &    0.3092072  & $\gamma=1.5$                          \\
\bf-3.5 & \bf0.5164954  &                                       \\
   -4.5 &    0.8408343  & $v_c= 1.0670662 \pm i 0.7483516$      \\
\bf-5.5 & \bf1.6284527  & ($\rightarrow$ singlet)               \\
   -6.5 &    2.2989503  &                                       \\
   -7.5 &    2.6116381  & $I_0=0$                               \\
   -8.5 &    2.8192199  &                                       \\
   -9.5 &    2.9854038  & $\Sigma=-1$                           \\
  -10.5 &    3.1355551  &                                       \\ \hline
% data6
    0.5 &    0.0407303  &                                       \\
   -0.5 &    0.1421697  & $\gamma=1$                            \\
   -1.5 &    0.2594036  &                                       \\
   -2.5 &    0.4157572  & $v_c= 1.3736073 \pm i 0.6398024 $     \\
   -3.5 &    0.6965562  & ($\rightarrow$ triplet)               \\
   -4.5 &    2.4190131  &                                       \\
\bf-5.5 & \bf2.7023676  & $I_0=0$                               \\
   -6.5 &    2.8629747  &                                       \\
   -7.5 &    2.9825513  & $\Sigma=+1$                           \\
\bf-8.5 & \bf3.0850192  &                                       \\ \hline
\end{tabular}
\end{tabular}
}
\vspace{0.5cm}
\caption{
\label{solutions}
Some solutions of the BAE (\ref{BAe}) of the $N=20$ lattice with 8
real rapidities, 2 holes and one closed pair. In the first two columns the
real rapidities $v$, the holes $\theta$ (emphasized by bold face) and the
corresponding quantum numbers $I$ 
can be found. The third column shows the parameter of the anisotropy $\gamma$,
the rapidities of the closed pair $v_c$ and the value of $I_0$ and $\Sigma$.
}
\end{table}

%%%%%%%%%%%%%%%%%%%%%%%%%%%%%%%%%%5

\end{document}